\documentclass{article}
\usepackage{spconf,amsmath,amssymb,graphicx,booktabs,xcolor,multirow,url}

\newcommand{\aag}{AAG}
\title{Accent Analogy Guidance: More Speaker Similarity\\ at Equal Accent in Cross-Lingual Voice Cloning}
\name{Yoomee Cho \qquad Jisun Lee}
\address{ESTsoft, Seoul, Republic of Korea\\
\{escym, jisun0424\}@estsoft.com}

\begin{document}
\setlength{\emergencystretch}{2em}
\maketitle

\begin{abstract}
In cross-lingual zero-shot text-to-speech, the accent of the reference leaks into the target speech.
We propose accent analogy guidance (\aag{}), a training-free sampler term that subtracts an accent direction estimated from the model's own predictions for one synthetic voice rendered in both languages, so the voice cancels and only the accent remains.
By a blind LLM accent judge on real dubbing data, reweighting classifier-free guidance between reference and text, and its variants, stay near one identity--accent trade-off curve; we score a method by its speaker similarity above that curve at equal accent ($\Delta$SIM).
Across four open TTS models \aag{} lies above the curve: on OmniVoice $\Delta$SIM is $+0.11$ to $+0.27$ on three test sets (accent 3.51 to 4.28 on a 1--5 scale at speaker similarity 0.29, where reweighting keeps 0.02); MaskGCT and CosyVoice 2 also lie above their curves, and on F5-TTS it is more native than any reweighting setting.
An LLM-free language-ID measure and a twelve-listener panel agree.
A premise test and the reach of a model's own curve indicate in advance whether and roughly how much \aag{} can gain, predicting the one model where it gains nothing (X-Voice).
\end{abstract}

\begin{keywords}
zero-shot TTS, cross-lingual voice cloning, accent, classifier-free guidance, dubbing
\end{keywords}

\section{Introduction}
\label{sec:intro}
Automatic dubbing re-voices translated lines in the original speaker's voice, which zero-shot TTS clones from a short reference, usually the speaker's own source-language line \cite{zhu2026omnivoice,wang2025maskgct,chen2025f5tts}.
When reference and target differ in language, the model also copies the reference's pronunciation, and the dub sounds foreign-accented \cite{zhang2023vallex,liu2026xvoice}.
The common inference-time remedy manipulates classifier-free guidance (CFG) \cite{ho2022cfg}: separate weights for reference and text \cite{jiang2025megatts3,yang2024dualspeech}, weights that switch across steps \cite{selectivecfg2025}, residual decompositions \cite{shi2026jrr}, or language-tag contrast \cite{lcg2026}. Training-time remedies add language identifiers with decoupled guidance \cite{liu2026xvoice} or separately learned accent representations \cite{crossaccent2026,lertpetchpun2026accentvector}, or, as in CosyVoice 2's cross-lingual mode, keep the reference out of the stage that chooses pronunciation \cite{du2024cosyvoice2}.
Training-time remedies need retraining; inference-time remedies rescale the same reference evidence, so accent and speaker move together (Sec.~\ref{sec:curve}). Results are usually reported as single operating points, without intervals, which leaves the identity cost of less accent unreported.

We make three contributions.
(i) We propose \aag{}, which contrasts the model's predictions for proxy voices rendered in both languages to estimate the accent direction with the speaker held fixed (Sec.~\ref{sec:method}).
(ii) We evaluate accent control by the speaker similarity it keeps at a given accent, relative to the model's own reweighting curve, and find that, by a blind judge, seven inference-time variants lie less than 0.05 above that curve across three sets, against 0.11--0.30 for \aag{} (Sec.~\ref{sec:curve}).
(iii) A premise test and the reach of a model's own curve indicate, before it is run, whether and roughly how much an inference-time fix can gain; we test this on five models, four on public test sets and one where it anticipates no effect (Sec.~\ref{sec:results}).

\section{Guidance and the identity--accent curve}
\label{sec:curve}
\begin{figure*}[t]\centering
\includegraphics[width=\textwidth]{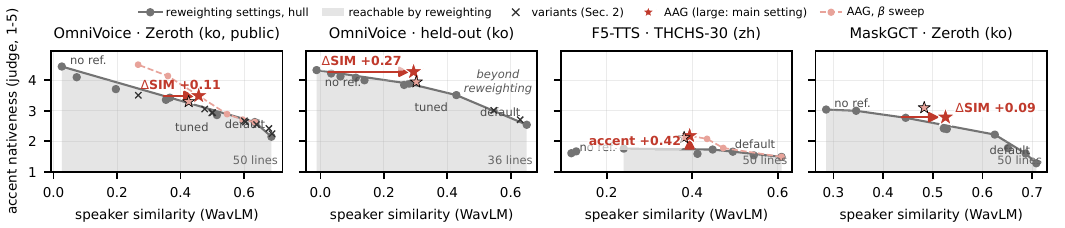}
\vspace{-6mm}
\caption{The identity--accent plane for four model--set pairs. Grey: the model's reweighting settings (dual weights, plus the no-reference end point) and their upper convex hull (shaded: reachable by reweighting); black crosses: the Sec.~\ref{sec:curve} variants; red stars: \aag{} (large: main setting); red dashes: its $\beta$ sweep. Axes: WavLM SIM to the real source-language line ($x$) and judge accent (1--5, $y$); $\Delta$SIM is the horizontal gap to the hull.}
\label{fig:plane}
\vspace{-3mm}
\end{figure*}
Let $x$ be the target speech, $c$ the reference and $y$ the target text.
With unconditional, text-only and full predictions $\ell_u$, $\ell_t$, $\ell_c$ (log-probabilities, embeddings or velocities), dual-weight guidance \cite{jiang2025megatts3} samples from
\begin{equation}
\ell = \ell_u + a_t\,(\ell_t-\ell_u) + a_s\,(\ell_c-\ell_t),
\label{eq:dual}
\end{equation}
where $\ell_t-\ell_u$ is the evidence the text adds and $\ell_c-\ell_t = \log p(c\mid x,y) + \text{const}$ the evidence the reference adds (for velocities, its score).
As a working model (not a derivation), let it factorise over the reference's speaker $s$ and accent $k_{\text{src}}$ (a product of experts \cite{liu2022composable}): $\log p(c\mid x,y) = \log p(s\mid x) + \log p(k_{\text{src}}\mid x) + \text{const}$. Then Eq.~\eqref{eq:dual} raises both factors to the same power $a_s$; weights, schedules or per-token reweighting of the same residuals scale both together at every position, so we expect them to trade speaker for accent along one curve (tested below).
The curve is the upper convex hull of reweighting settings $(a_s,a_t)$ and the no-reference end point, a conservative baseline since mixing two settings line by line reaches any point between them.
We score a method by $\Delta$SIM, its speaker similarity (SIM) minus the curve's at equal accent (Fig.~\ref{fig:plane}), and by $\Delta$accent, its accent gain at equal SIM. The 95\,\% CIs come from a speaker bootstrap that redraws the curve; paired per-line differences resample lines.
We test seven variants: four reweightings of the same residuals (a weight switch across steps, reference-free early steps, a per-token floor on the expected text evidence, and a Fisher-metric projection of the reference residual) and three added directions (guidance toward the line's own target-language draft, language-tag contrast after \cite{lcg2026}, 6 settings, and accent-label instruction contrast, 2). All variants were run on OmniVoice. By the judge, none lies 0.05 or more above the hull on the 49-line dubbing pilot or the public Zeroth-Korean and THCHS-30 sets (Sec.~\ref{sec:exp}); 5 of 17 variant--set pairs (Zeroth, THCHS) are significantly above it, all by under 0.05 and at low accent. \aag{} (Sec.~\ref{sec:method}) scores $+0.30$, $+0.11$ and $+0.15$ there (held-out $+0.27$) and beats the best variant on each set by $+0.09$ to $+0.28$ (95\,\% CIs above 0).

\section{Accent analogy guidance}
\label{sec:method}
\noindent\textbf{Direction by analogy.}
Take a proxy voice $v$ with two renditions that differ only in language: $c_v^{\text{src}}$ (the voice speaking the source language) and $c_v^{\text{tgt}}$ (the same voice speaking the target language).
Under the factorisation the speaker cancels in the difference of their predictions,
\begin{equation}
d_v = \ell_{c_v^{\text{src}}} - \ell_{c_v^{\text{tgt}}} = \log p(k_{\text{src}}\mid x) - \log p(k_{\text{tgt}}\mid x),
\end{equation}
so the difference of the model's own predictions gives the accent direction. Averaging $d_v$ over $M$ proxies, \aag{} with settings $(a_s,a_t,\beta)$ samples from
\begin{equation}
\ell_{\aag} = \ell_u + a_t(\ell_t-\ell_u) + a_s(\ell_c-\ell_t) - \tfrac{\beta}{M}\textstyle\sum_{m} d_{v_m} .
\label{eq:aag}
\end{equation}
With $\beta = a_s$ this is guidance toward the counterfactual reference ``same speaker, target-language accent'', $a_s[\log p(s\mid x)+\log p(k_{\text{tgt}}\mid x)]$, which exists as a direction but not as audio. Semantic guidance \cite{brack2023sega}, negation in composable diffusion \cite{liu2022composable} and contrastive steering \cite{rimsky2024caa} add or subtract prompt- or activation-derived directions; here the subtracted direction is a difference of two conditions that share a voice, so identity cancels.

\begin{table}[t]\centering\footnotesize
\caption{When does \aag{} help? Premise: accent lost when English is cloned from a proxy's source-language rendition instead of its English one ($^\S$renditions by OmniVoice / X-Voice). Reach: SIM the hull keeps at accent 4.0, on sets whose hull gets there (F5-TTS, CosyVoice 2: their most native accent). \aag{}: $\Delta$SIM of the main setting across test sets (MaskGCT: both of its settings; X-Voice: pilot); $^\dagger$accent gain over the most native curve setting.}
\label{tab:scope}
\setlength{\tabcolsep}{3pt}
\begin{tabular}{@{}lccc@{}}\toprule
Model (pair) & premise & reach & \aag{} \\\midrule
OmniVoice (ko$\to$en) & 1.33 (zh 0.61) & 0.18--0.26 & $+0.11$--$0.27$ \\
MaskGCT (ko$\to$en) & 1.95 & 0.32--0.38 & $+0.07$--$0.20$ \\
F5-TTS (zh$\to$en) & 0.48 & (max 1.8) & $+0.42^\dagger$/$+0.36^\dagger$ \\
X-Voice (ko$\to$en) & 0.00/0.33$^\S$ & 0.29 & $-0.05$ \\
CosyVoice 2 (ko$\to$en) & 0.86 & (max 2.9) & $+0.07$; $+0.37^\dagger$ \\\bottomrule
\end{tabular}
\vspace{-5mm}
\end{table}

\begin{table*}[t]\centering\footnotesize
\caption{Main results (line means, 2 seeds). ``tuned'': $(2,5)$; ``equal-accent rew.'': the sampled setting closest in accent to \aag{}; ``alt.'': alternative setting, fixed in advance. ``curve'': the SIM the hull keeps at the row's accent ($^\dagger$: at its most native point); $\Delta$SIM $=$ SIM $-$ curve on the judge axis (95\,\% speaker CIs; bold: CI above 0) and the language-ID axis ($^*$CI excludes 0). LID: logit P(en).}
\label{tab:main}
\setlength{\tabcolsep}{4pt}
\begin{tabular}{llcccccccc}\toprule
 Model, set & setting & accent$\uparrow$ & LID$\uparrow$ & SIM$\uparrow$ & curve & WER\,\%$\downarrow$ & UTMOS$\uparrow$ & $\Delta$SIM & $\Delta$SIM, LID \\\midrule
\multirow{4}{*}{\shortstack[l]{OmniVoice\\\scriptsize held-out ko, 36}} & default & 2.54 & 1.43 & 0.651 &  & 5.7 & 3.25 &  &  \\
 & tuned & 3.51 & 3.85 & 0.429 &  & 4.1 & 3.73 &  &  \\
 & equal-accent rew. & 4.22 & 5.00 & 0.034 &  & 4.9 & 4.08 &  &  \\
 & \textbf{\aag{}} & 4.28 & 5.03 & 0.294 & 0.02 & 5.1 & 3.72 & $\mathbf{+0.27}$ {\scriptsize $[+0.02,+0.34]$} & $+0.18$$^*$ \\
\midrule
\multirow{4}{*}{\shortstack[l]{OmniVoice\\\scriptsize Zeroth ko, 50}} & default & 2.15 & 2.54 & 0.686 &  & 5.3 & 3.16 &  &  \\
 & tuned & 2.86 & 4.26 & 0.514 &  & 4.1 & 3.78 &  &  \\
 & equal-accent rew. & 3.43 & 4.71 & 0.366 &  & 3.3 & 4.10 &  &  \\
 & \textbf{\aag{}} & 3.49 & 4.88 & 0.457 & 0.35 & 3.8 & 3.58 & $\mathbf{+0.11}$ {\scriptsize $[+0.04,+0.16]$} & $+0.09$$^*$ \\
\midrule
\multirow{4}{*}{\shortstack[l]{OmniVoice\\\scriptsize THCHS zh, 50}} & default & 1.17 & 3.35 & 0.753 &  & 9.3 & 3.94 &  &  \\
 & tuned & 1.93 & 5.22 & 0.556 &  & 9.2 & 4.12 &  &  \\
 & equal-accent rew. & 2.41 & 5.65 & 0.365 &  & 10.2 & 4.22 &  &  \\
 & \textbf{\aag{}} & 2.32 & 5.08 & 0.569 & 0.42 & 8.0 & 3.94 & $\mathbf{+0.15}$ {\scriptsize $[+0.06,+0.20]$} & $-0.00$ \\
\midrule
\multirow{4}{*}{\shortstack[l]{F5-TTS\\\scriptsize THCHS zh, 50}} & default & 1.49 & 9.40 & 0.609 &  & 10.4 & 3.11 &  &  \\
 & equal-accent rew. & 1.76 & 7.89 & 0.240 &  & 9.1 & 3.33 &  &  \\
 & \textbf{\aag{}} & 2.18 & 5.85 & 0.394 & 0.24$^\dagger$ & 26.4 & 2.50 & $+0.15$ {\scriptsize $[-0.08,+0.28]$}$^\dagger$ & $-0.21$$^*$ \\
 & \textbf{\aag{} (alt.)} & 2.12 & 7.18 & 0.381 & 0.24$^\dagger$ & 9.7 & 3.16 & $+0.14$ {\scriptsize $[-0.11,+0.26]$}$^\dagger$ & $-0.23$$^*$ \\
\midrule
\multirow{4}{*}{\shortstack[l]{MaskGCT\\\scriptsize Zeroth ko, 50}} & default & 1.28 & 0.88 & 0.708 &  & 24.7 & 2.15 &  &  \\
 & equal-accent rew. & 2.77 & 6.56 & 0.445 &  & 5.5 & 2.78 &  &  \\
 & \textbf{\aag{}} & 2.79 & 6.53 & 0.526 & 0.44 & 11.0 & 2.39 & $+0.09$ {\scriptsize $[-0.02,+0.20]$} & $+0.08$$^*$ \\
 & \textbf{\aag{} (alt.)} & 3.10 & 6.87 & 0.483 & 0.29$^\dagger$ & 6.1 & 2.72 & $\mathbf{+0.20}$ {\scriptsize $[+0.03,+0.22]$}$^\dagger$ & $+0.16$$^*$ \\
\bottomrule
\end{tabular}

\vspace{-5mm}
\end{table*}

\noindent\textbf{Premise test.}
\aag{} can only remove an accent the reference's language induces; Table~\ref{tab:scope} tests this per model by cloning from $c_v^{\text{tgt}}$ and from $c_v^{\text{src}}$.

\noindent\textbf{Proxy set.}
$c_v^{\text{tgt}}$ is a native-accented sample (here a no-reference OmniVoice generation that the judge rates fully native); $c_v^{\text{src}}$ is the same voice cloned into the source language by the model under test (for X-Voice, by OmniVoice). One set per language pair and gender ($M{=}2$) serves every speaker; no test speaker is used. For OmniVoice they are the first two English voices per gender; for the other models, the two per gender whose renditions moved the accent most in the premise test (a property of the proxy, measured without test data).

\noindent\textbf{Choosing $\beta$.}
On the design pilot (Sec.~\ref{sec:exp}) we chose $\beta{=}5$ at $a_s{=}3$, above the counterfactual reading's $\beta{=}a_s$. A post-hoc account: proxy renditions, clones of a target-language voice, have less source accent than real speakers, so $\beta\approx r\,a_s$ with $r>1$ restores the counterfactual reading. Its two prospective uses (the primary settings of MaskGCT and CosyVoice 2) were not significant; $\beta$ still needs a small per-model sweep.

\noindent\textbf{Instantiation and cost.}
Eq.~\eqref{eq:aag} acts on log-probabilities (OmniVoice, CosyVoice 2), pre-softmax embeddings (MaskGCT's text-to-semantic stage) or velocities over the target frames (F5-TTS, X-Voice); each proxy adds two network evaluations per step (3.3 vs.\ 2.0\,s per OmniVoice line on an RTX 3090).

\section{Experimental setup}
\label{sec:exp}
\noindent\textbf{Models.} OmniVoice \cite{zhu2026omnivoice} (masked diffusion over codec tokens), MaskGCT \cite{wang2025maskgct} (\aag{} in the text-to-semantic stage; the acoustic stage keeps the real speaker's prompt), F5-TTS \cite{chen2025f5tts} (flow matching; the prompt transcript is part of its text input), X-Voice \cite{liu2026xvoice} (flow matching with language-ID injection and decoupled CFG) and CosyVoice 2 \cite{du2024cosyvoice2} (a token LM, where \aag{} acts, with no CFG of its own: $(1,1)$; then a flow decoder; zero-shot mode), all public checkpoints. Default CFG scales correspond to $(a_s,a_t)=(3,3)$ (OmniVoice, F5-TTS) and $(3.5,3.5)$ (MaskGCT, X-Voice); for OmniVoice we also report $(2,5)$, the operating point chosen for dubbing before this work (``tuned''). With $\beta{=}0$ our samplers reproduce the originals token for token.

\noindent\textbf{Data.} A pilot of 49 dubbing lines from 15 real videos in 7 source languages served OmniVoice's design choices; its settings were then fixed, and every test set below was run once with them, without further tuning. The held-out Korean set has 36 lines from 9 of the same speakers. The public sets are Zeroth-Korean \cite{zerothkorean} (50 lines, 10 speakers) and THCHS-30 \cite{wang2015thchs30} (50 dev lines for the Chinese settings and 50 test lines, 10 speakers). Targets are Gemini 2.5 translations; we use two seeds (pilot: up to three) unless stated.

\noindent\textbf{Metrics.} We measure accent with a blind Gemini 2.5 Pro judge (1--5, 5 = native). On real speech it tracks human experts on 154 sentences by Mandarin-L1 adults (Spearman 0.74 $[0.65,0.81]$, speechocean762 \cite{zhang2021speechocean762}) and separates these from 80 lines by 40 native LibriSpeech readers \cite{panayotov2015librispeech} (AUC 0.97; natives 4.42, L2 sentences the experts rate 9/10: 2.86 $[2.62,3.17]$, those under 7: the floor); repeat ratings agree ($\kappa$ 0.92). Without an LLM, Whisper large-v3's \cite{radford2023whisper} language-ID logit, logit\,P(en), separates the same two groups (AUC 0.995) and follows the experts (Spearman 0.59). Speaker similarity is WavLM verification as in Seed-TTS-eval \cite{chen2022wavlm,anastassiou2024seedtts} (one real speaker: 0.81), complemented by closed-set speaker identification. Intelligibility is Whisper WER and naturalness UTMOS \cite{saeki2022utmos}.

\section{Results}
\label{sec:results}
\noindent Samples: {\small\url{https://yoomee-cho.github.io/accent-analogy-guidance/}}.
\noindent Across the four models where the premise holds, \aag{} lies above the model's own reweighting curve (OmniVoice on three test sets, MaskGCT, CosyVoice 2) or beyond its accent range (F5-TTS); on X-Voice, where the premise fails, it gains nothing (Tables~\ref{tab:scope} and \ref{tab:main}).

\noindent\textbf{OmniVoice} (Table~\ref{tab:main}).
On public Zeroth-Korean, the fully independent test, $(3,5,5)$ raises the accent from 2.86 (the tuned setting; the level of the best-rated L2 readers above) to 3.49 at SIM 0.457 (tuned: 0.514): $\Delta$SIM $+0.11$ $[+0.04,+0.16]$, $\Delta$accent $+0.33$ $[+0.12,+0.46]$.
On the held-out dubbing set the effect is larger: the accent rises from 3.51 to 4.28 (paired $+0.76$ $[+0.46,+1.07]$), within 0.05 of a voice generated with no reference at all (4.33), while SIM stays at 0.294 (tuned: 0.429) where the hull keeps only 0.02: $\Delta$SIM $+0.27$ $[+0.02,+0.34]$.
In other words, on real dubbing lines \aag{} delivers a near-native accent and keeps two thirds of the tuned setting's speaker similarity, where reweighting keeps none.
With Chinese as the source (THCHS) it raises the accent by $+0.39$ $[+0.21,+0.59]$ at unchanged SIM (0.57 vs.\ 0.56; $\Delta$SIM $+0.15$), and for Spanish targets (Zeroth) from 4.42 to 4.81 ($+0.39$ $[+0.21,+0.58]$; $\Delta$SIM $+0.08$ $[-0.02,+0.22]$, n.s.; the alternative $(2,5,2)$: $+0.13$ $[+0.02,+0.22]$): three language pairs, one setting, no retuning.
Identity is preserved: in closed-set identification among all 41 speakers (chance 2.4\,\%), \aag{} keeps most of the tuned setting's identifiability (Zeroth 95 vs.\ 97\,\%, THCHS 99 vs.\ 90\,\%, held-out 69 vs.\ 81\,\%), whereas reweighting pushed to the same accent drops to 81, 71 and, on held-out, 4\,\%.
The cost is small: WER stays within one point of the tuned setting's (3.8 vs.\ 4.1, 5.1 vs.\ 4.1, 8.0 vs.\ 9.2\,\%), UTMOS within 0.20, and no model is trained.
Finally, $\beta$ is a single knob that spans the whole trade-off (Fig.~\ref{fig:plane}, Zeroth): \aag{} leaves the hull from $\beta{=}5$, and at $\beta{=}8$ it is as native as the no-reference voice (4.51 vs.\ 4.45) at SIM 0.27 instead of 0.03.

\noindent\textbf{Other models} (Table~\ref{tab:scope}).
The same recipe transfers to three other public models without retraining, and the size of the gain follows the diagnostic.
MaskGCT's acoustic stage re-applies the speaker prompt's timbre, so reweighting already keeps SIM 0.27--0.35 at its most native settings and \aag{} has less to gain.
Still, on Zeroth its alternative setting $(1.5,5,2)$ is more native than every reweighting setting (3.10 vs.\ at most 3.04) while keeping SIM 0.48 instead of 0.29, $+0.20$ $[+0.03,+0.22]$ above the hull; the primary setting gains $+0.09$ $[-0.02,+0.20]$ (held-out: $+0.07$, n.s.).
F5-TTS is the case where reweighting barely moves the accent (at most 1.76 across the grid), and \aag{} is more native than every reweighting setting: by $+0.36$ $[+0.14,+0.57]$ for $(2,5,2)$ at SIM 0.38 vs.\ 0.24 with similar WER and UTMOS, and by $+0.42$ $[+0.18,+0.68]$ for $(3,3,5)$ at a cost in WER (26 vs.\ 9\,\%) and UTMOS (2.50 vs.\ 3.33); language ID, saturated (all P(en) $>$ 0.99), cannot confirm this.
On CosyVoice 2, whose premise holds only weakly (0.86 $[0.10,1.57]$), the primary setting $(1,1,1.67)$ is above the hull by $+0.07$ $[-0.03,+0.08]$ (language ID $+0.07$, CI above 0) and more native than any setting on its curve, even its own cross-lingual mode ($+0.37$ $[+0.11,+0.64]$; SIM 0.53 vs.\ 0.46).
X-Voice is the predicted exception: \aag{} does not help ($-0.05$ $[-0.20,+0.09]$, pilot, 1 seed) because with the voice fixed its accent barely depends on the reference's language, so there is nothing to remove.
Premise alone does not rank the eight outcomes (Spearman $+0.07$); with reach, the SIM reweighting already keeps, it accounts for them (Table~\ref{tab:scope}).

\noindent\textbf{Controls.}
Three checks tie the gain to the accent direction itself.
Proxies rendered in the wrong language (Spanish for Korean lines, same weights) remove it on the pilot (accent 3.49 vs.\ 4.33, the tuned setting 3.52; $\Delta$SIM $+0.03$, n.s.): the effect is language-specific, not a general pull toward native voices.
\aag{} is not a per-token cut of $a_s$ either: the part of its direction parallel to $\ell_c-\ell_t$ keeps at most 37\,\% of the gain (Zeroth $+0.01$, held-out $+0.10$).
And proxy identity cancels on average: leakage (SIM to the proxies used minus that to unused ones) is $+0.02$ to $+0.06$ with two proxies ($+0.17$ with one).
The gain does depend on which proxies are used: two other sets from the same 11 voices match it on held-out ($+0.26$, $+0.23$) but give about half of it on Zeroth ($+0.05$, $+0.06$).

\noindent\textbf{Language ID.}
The result does not rest on the LLM judge alone.
Whisper's language ID follows the judge across the 45 pilot settings (Spearman 0.97) and also puts \aag{} above the hull on the pilot ($+0.20$ $[+0.08,+0.33]$), Zeroth and held-out; it does not on THCHS (Table~\ref{tab:main}), for Spanish targets or with wrong-language proxies (all n.s.).

\noindent\textbf{Listening panel.}
Human listeners confirm both halves of the claim.
Twelve raters (eleven Korean-L1, one Spanish-L1; six near-native or fluent in English; none excluded by the catch rule) compared \aag{} with the tuned setting and with equal-accent reweighting on 20 Zeroth lines, with hypotheses fixed before the test (CIs: a two-way cluster bootstrap over raters and lines).
They heard \aag{} as more native than the tuned setting in 85\,\% of decided answers $[64,99]$, and as closer to the reference speaker than equal-accent reweighting in 85\,\% $[69,97]$; the six most proficient raters agree (84 and 83\,\%).
The trade-off is audible: against the tuned setting, the tuned voice was judged closer in 77\,\% $[53,94]$; \aag{} offers a better trade-off than reweighting, not a free one.
On naturalness (secondary) they preferred \aag{} to equal-accent reweighting, 73\,\% $[52,92]$, the opposite of UTMOS's ranking.

\section{Conclusion}
\label{sec:conclusion}
Inference-time accent control has been confined to one curve: by the judge, reweighting CFG trades accent for speaker similarity, and none of seven variants leaves that curve by more than 0.05. \aag{} leaves it. Contrasting the model's own predictions for one synthetic voice in two languages isolates the accent direction and removes it, without training, in token, logit and flow samplers alike. On OmniVoice this yields near-native dubbing (accent 4.28 vs.\ 4.33 for a reference-free voice) while keeping two thirds of the speaker similarity that reweighting gives up, with the speaker still identifiable (69--99\,\% across sets) and WER within one point; twelve listeners confirm both the accent and the identity gain. It transfers without retraining to MaskGCT, F5-TTS and CosyVoice 2, where a setting is more native than the whole reweighting curve, and a premise test with the curve's reach predicts in advance how much a model can gain, including the one that gains nothing (X-Voice). Its main limitations are an LLM judge as the primary measure (language ID disagrees on THCHS, Spanish and F5), sensitivity to the proxy set, and a per-model $\beta$ sweep. Code and proxy sets will be released.

\bibliographystyle{IEEEbib}
\bibliography{refs}

\vspace{3mm}\noindent\textbf{Acknowledgements.} We thank the twelve volunteers who took the listening test.

\vspace{3mm}\noindent\textbf{Compliance with Ethical Standards.}
The listening test collects A/B preferences from adult volunteers who agreed to take part; names were used only to track participation and are not kept with the results. The held-out dubbing lines come from a commercial dubbing service whose privacy policy allows user inputs to be used to improve the service and to train models; they were used for evaluation only, and no audio or transcript from them is released. All released samples come from public corpora.
\end{document}